\documentclass[12pt, aps,prl,onecolumn,superscriptaddress, showkeys]{revtex4-2}

\usepackage{amsmath, physics}
\usepackage{amssymb}
\usepackage{bm}

\usepackage{geometry}
\usepackage{graphicx} 
\usepackage{mathrsfs}

\usepackage{tikz}
\usetikzlibrary{3d,calc}
\usepackage{tikz-3dplot}

\newcommand{\vv}[1]{\bm{#1}} 

\date{\today}

\begin{document}

\title{Deriving the von Neumann equation from the Majorana--Bloch equation for arbitrary spin in any state}

\author{Lihong V. Wang}

\email{LVW@caltech.edu}

\affiliation{Caltech Optical Imaging Laboratory, Andrew and Peggy Cherng Department of Medical Engineering, Department of Electrical Engineering, California Institute of Technology, Pasadena, CA 91125, USA}

\begin{abstract}
After publishing the derivation from the classical Bloch equation to the quantum von Neumann equation to the Schrödinger--Pauli equation for spin-\(\tfrac{1}{2}\), we proposed renaming the Bloch equation to the Majorana--Bloch equation because Majorana's work predated Bloch's in the presentation of the Bloch equation by 14 years. Here, we first generalize our previous derivation to higher spins or angular momenta in coherent pure states. Using the polynomial representation of the coherent-state projector, we derive an invertible mapping from the Majorana--Bloch equation to the von Neumann equation, establishing a one-to-one correspondence between these two formalisms. Application of the Ehrenfest theorem also shows that expectation values in these states reproduce the classical equation of motion as expected. Then, we obtain arbitrary spin-$s$ states by symmetrizing tensor products of spin-$\tfrac{1}{2}$ primitives, in accordance with the Majorana construction or the Schur--Weyl duality.

\end{abstract}


\maketitle

\section{Introduction}

Since our original publication on the derivation from the classical Bloch equation to the quantum von Neumann equation to the Schrödinger--Pauli equation for spin-\(\tfrac{1}{2}\) \cite{Wang2022}, we added an appendix to detail the derivation of the Bloch or Majorana--Bloch equation \cite{Wang2024}. Because Majorana \cite{majorana1932atomi, majorana2020oriented} introduced the formulation in 1932, now known as the Bloch equation, prior to Bloch’s 1946 publication \cite{Bloch1946nuclear}, we suggest renaming it the Majorana--Bloch equation \cite{Wang2024}. Building on our previous work~\cite{Wang2022, Wang2024, Wang2022SG, Wang2023SG}, we extend the derivation from the Majorana–Bloch equation to the von Neumann equation for arbitrary spins or angular momenta, with a focus on the pure coherent-state manifold. The inverse derivation holds, whether or not expectation values are taken. Further, we construct general spin states by symmetrizing tensor products of spin-$\tfrac{1}{2}$ primitives \cite{sakurai2020modern}, following the Majorana construction or the Schur--Weyl duality \cite{majorana1932atomi, majorana2020oriented}. In what follows, the term ``spin'' is used to refer to either spin or orbital angular momentum, as appropriate.

The work here is limited to Hamiltonians linear in spin, uniform in space, and time-dependent. The bijection holds on the coherent-state manifold. Note that the mapping does not extend to non-linear spin Hamiltonians. 

\section{Derivation from Majorana--Bloch equation to von Neumann equation for arbitrary spin in coherent state} \label{sec:MBE-vNE}

\begin{figure}
    \centering

\resizebox{0.6\linewidth}{!}{ 
\tdplotsetmaincoords{70}{110}
\def\thetaVal{45}
\def\phiVal{45}

\begin{tikzpicture}[tdplot_main_coords,scale=3,>=stealth]

  \draw[->] (0,0,0)--(1.2,0,0)   node[anchor= west]{$x$};
  \draw[->] (0,0,0)--(0,1.2,0)   node[anchor= west]{$y$};
  \draw[->] (0,0,0)--(0,0,1.2)   node[anchor=west]     {$z$};

  \draw[->,thick,blue] (0,0,0)--(0,0,1) node[anchor=west]{$\vv B$};

  \tdplotsetcoord{psi}{2}{\phiVal}{\thetaVal}
  \draw[->,thick,red] (0,0,0)--(psi) node[anchor= west]{$\vv{n}_0, |s\rangle_{\vv{n}_0}$};

  \coordinate (proj) at
    ({2*sin(\thetaVal)*cos(\phiVal)},
     {2*sin(\thetaVal)*sin(\phiVal)},0);
  \draw[dashed] (psi)--(proj)--(0,0,0);

  \tdplotdrawarc[->,dashed]{(0,0,0)}{0.3}{0}{\phiVal}{anchor=north}{};
\node at (.2,0.15,0) [anchor=north east]{$\phi$};

  \draw[dashed,->]
plot[domain=0:\thetaVal,samples=50,variable=\t]
      ({0.3*sin(\t)*cos(\phiVal)},
       {0.3*sin(\t)*sin(\phiVal)},
       {0.3*cos(\t)})
    node[pos=0.3,anchor=south west]{};
    \node at (.2,0.25,0.55) [anchor=north east]{$\theta$};
    
\end{tikzpicture}
}

    \caption{Illustration of the coordinates and variables.}
    \label{fig:xyz}
\end{figure}
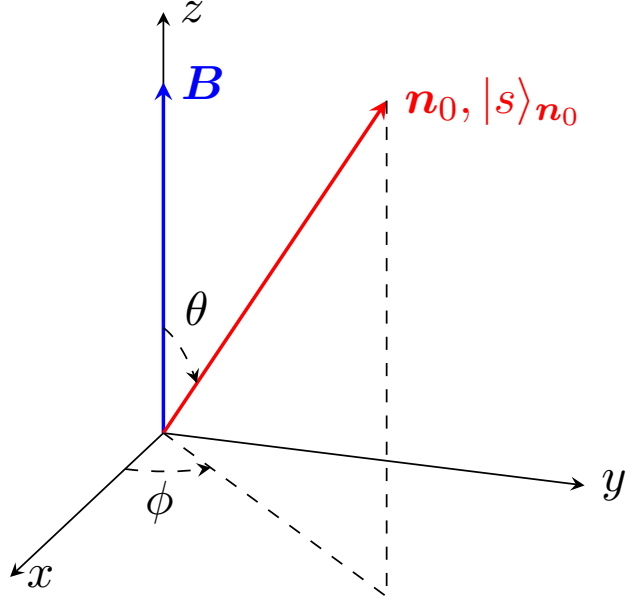

We begin with the classical Majorana--Bloch equation,
\begin{equation}
    \frac{d\vv S_{\mathrm{cl}}}{dt} = 
    \gamma \, \vv S_{\mathrm{cl}} \times \vv{B},
\end{equation}
where \(\vv S_{\mathrm{cl}}\) denotes the classical spin pseudovector, \(\gamma\) the gyromagnetic ratio, \(\vv{B}\) the magnetic flux density pseudovector, and \(t\) time. We define
\begin{equation}
\vv{n}_0
\equiv
\frac{\vv S_{\mathrm{cl}}}{\hbar\,s},
\label{eq:n_hat_def}
\end{equation}
where $\hbar$ is the reduced Planck constant, $s$ is the spin angular momentum quantum number, and $\vv{n}_0$ is a unit vector (i.e., $|\vv{n}_0|=1$, Fig. \ref{fig:xyz}). Substitution gives
\begin{equation}
\frac{d\vv{n}_0}{dt}
=\gamma\,\vv{n}_0\times\vv B.
\label{eq:MBE_n}
\end{equation}
Multiplying both sides from the right by the spin‑\(s\) angular momentum operator pseudovector \(\hat{\vv{S}}  = (\hat{S}_x, \hat{S}_y, \hat{S}_z)\) yields
\begin{equation}
\frac{d}{dt}\bigl(\vv{n}_0\cdot\hat{\vv S}\,\bigr)
=\gamma\,\bigl(\vv{n}_0\times\vv B\,\bigr)\cdot\hat{\vv S}.
\end{equation}
From the commutator identity valid for any vectors \(\vv{a}, \vv{b} \in \mathbb{R}^3\) (see Appendix \ref{sec:CommCross}),
\begin{equation}
\bigl(\vv a\times\vv b\,\bigr)\cdot\hat{\vv S} 
= \frac{1}{i\hbar}
\bigl[\vv a\cdot\hat{\vv S},\;\vv b\cdot\hat{\vv S}\,\bigr],
\end{equation}
which holds for any spin representation, we obtain
\begin{equation}
\frac{d}{dt}\bigl(\vv{n}_0\cdot\hat{\vv S}\,\bigr)
=\frac{\gamma}{i\hbar}\,\bigl[\vv{n}_0\cdot\hat{\vv S}, \vv B\cdot\hat{\vv S}\,\bigr].
\label{eq:dSdt}
\end{equation}

We define the Zeeman-type Hamiltonian
\begin{equation}
\hat{H}\equiv-\,\gamma\;\vv B\cdot\hat{\vv S}
\end{equation}
and the dimensionless projection of \(\hat{\vv S}\) along \(\vv{n}_0\)
\begin{equation}
\hat{X}_{\vv{n}_0} \equiv \frac{1}{\hbar}\,\vv{n}_0\cdot\hat{\vv{S}}.
\end{equation}
Consequently, Eq. \ref{eq:dSdt} becomes

\begin{equation}
    \frac{d\hat{X}_{\vv{n}_0}}{dt} =\frac{1}{i \hbar}\,[\hat{H},\,\hat{X}_{\vv{n}_0}]. \label{eq:dXdt}
\end{equation}

For the coherent state along \(\vv{n}_0\), we find its density operator (see Appendix \ref{sec:rho}),
\begin{equation}
{\rho}_{\vv{n}_0}
= \prod_{m'=-s}^{s-1}\frac{\hat{X}_{\vv{n}_0}-m' \hat{I}}{s-m'},
\label{eq:rho_poly1}
\end{equation}
where \(\hat{I}\) denotes the identity operator. We write Eq. (\ref{eq:rho_poly1}) with fixed coefficients $a_k$, which depend only on $s$, as
\begin{equation}
    {\rho}_{\vv{n}_0}
    =\sum_{k=0}^{2s}a_k \hat{X}_{\vv{n}_0}^{k},
    \label{eq:rho_n_poly}
\end{equation}
a polynomial of degree $2s$. Then,
\begin{equation}
\frac{d{\rho}_{\vv{n}_0}}{dt}=\sum_{k=0}^{2s}a_k\frac{d\hat{X}_{\vv{n}_0}^{k}}{dt}
=\sum_{k=0}^{2s}a_k\sum_{j=0}^{k-1}\hat{X}_{\vv{n}_0}^{j}\,\frac{d\hat{X}_{\vv{n}_0}}{dt}\,\hat{X}_{\vv{n}_0}^{k-1-j}.
\label{eq:drhodt}
\end{equation}
Inserting (\ref{eq:dXdt}) inside the sum produces
\begin{equation}
i \hbar\,\frac{d{\rho}_{\vv{n}_0}}{dt}
=\sum_{k=0}^{2s}a_{k}
  \sum_{j=0}^{k-1}\hat{X}_{\vv{n}_0}^{j}\,[\hat{H}, \hat{X}_{\vv{n}_0}]\,\hat{X}_{\vv{n}_0}^{\,k-1-j}.
\end{equation}
Using (see Appendix \ref{sec:CommExp})
\begin{equation}
\sum_{j=0}^{k-1}\hat{X}_{\vv{n}_0}^{j}\,[\hat{H}, \hat{X}_{\vv{n}_0}]\,\hat{X}_{\vv{n}_0}^{\,k-1-j}
=[\hat{H}, \hat{X}_{\vv{n}_0}^{k}],
\end{equation}
we obtain
\begin{equation}
i \hbar\,\frac{d{\rho}_{\vv{n}_0}}{dt}
=\sum_{k=0}^{2s}a_{k}\,[\hat{H}, \hat{X}_{\vv{n}_0}^{k}].
\end{equation}
Substituting Eq. \ref{eq:rho_n_poly} yields
\begin{equation}
i\hbar\,\frac{d{\rho}_{\vv{n}_0}}{dt}
=[\hat{H},{\rho}_{\vv{n}_0}],
\label{eq:vNEq}
\end{equation}
which is the space‑independent von Neumann equation for arbitrary spin \(s\) or any angular momentum. For the pure coherent‑state manifold, the derivation establishes an exact correspondence from the classical (real‑space) Majorana--Bloch dynamics to the quantum (Hilbert‑space) operator formalism. 

Since each step is reversible on the coherent-state manifold for linear Zeeman dynamics, the derivation is bijective. Hence, the von Neumann equation and the Majorana--Bloch equation are equivalent descriptions of the same rigid-precession dynamics on this manifold. The previous steps are reversible under the stated assumptions, which emphasizes that the reversibility holds only within the conditions specified and does not imply general reversibility beyond this setting.

\section{Expectation‑value derivation of the classical Majorana--Bloch equation in coherent state}
The von Neumann equation can also be recast into the Majorana--Bloch form by taking expectation values. First, we prove that the expectation value of the quantum spin operator $\hat{\vv S}$ under the coherent‐state density operator recovers the classical spin pseudovector:
\begin{equation}
\Tr\bigl(\rho_{\vv{n}_0}\,\hat{\vv S}\,\bigr)
= \vv S_{\mathrm{cl}}.
\label{eq:exp_S=S_cl}
\end{equation}
Note that
\begin{equation}
    \Tr\bigl({\rho}_{\vv{n}_0}\,\hat{\vv S}\,\bigr)
=\bigl(\Tr({\rho}_{\vv{n}_0}\,\hat{S}_x),\,\Tr({\rho}_{\vv{n}_0}\,\hat{S}_y),\,\Tr({\rho}_{\vv{n}_0}\,\hat{S}_z)\bigr).
\end{equation}

For brevity, we define
\begin{equation}
\lvert\vv{n}_0\rangle
\equiv
\lvert s\rangle_{\vv{n}_0}.
\label{eq:ket_nhat_def}
\end{equation}

From Eq. \ref{eq:rho_def} (see Appendix \ref{sec:rho}), we obtain
\begin{equation}
\Tr\bigl(\rho_{\vv{n}_0}\,\hat{\vv S}\,\bigr)
= \langle \vv{n}_0\rvert\,\hat{\vv S}\,\lvert \vv{n}_0\rangle.
\label{eq:Tr_rho_S}
\end{equation}
Eq. \ref{eq:X_eigen} gives
\begin{equation}
\bigl(\vv{n}_0\cdot\hat{\vv S}\,\bigr)\,\lvert \vv{n}_0\rangle
= \hbar\,s\,\lvert \vv{n}_0\rangle,
\label{eq:n_dot_S_eigen}
\end{equation}
leading to
\begin{equation}
    \langle \vv{n}_0\rvert\,\bigl(\vv{n}_0 \cdot \hat{\vv S}\,\bigr)\,\lvert \vv{n}_0\rangle
=\hbar\,s.
\label{eq:n_dot_S_exp}
\end{equation}

One can show that the perpendicular components vanish using ladder operators in the local coordinates aligned with \(\vv{n}_0\), 
\begin{equation}
\hat S_{\vv{n}_0,\pm}
=\hat S_{\vv{n}_0,x} \pm i\,\hat S_{\vv{n}_0,y},
\qquad
\hat S_{\vv{n}_0, z} \equiv \vv{n}_0\cdot\hat{\vv S}.
\end{equation}
For the highest-weight eigenstate $\lvert \vv{n}_0\rangle$,
\begin{equation}
\hat S_{\vv{n}_0,+}\,\lvert \vv{n}_0\rangle = 0,
\qquad
\langle \vv{n}_0|\,\hat S_{\vv{n}_0,-} = 0.
\end{equation}
Hence,
\begin{equation}
\langle \vv{n}_0|\,\hat S_{\vv{n}_0,x}\,\lvert \vv{n}_0\rangle
=\frac12\,\langle \vv{n}_0|(\hat S_{\vv{n}_0,+}+\hat S_{\vv{n}_0,-})\lvert \vv{n}_0\rangle=0,
\end{equation}
\begin{equation}
\langle \vv{n}_0|\,\hat S_{\vv{n}_0,y}\,\lvert \vv{n}_0\rangle
=\frac{1}{2i}\,\langle \vv{n}_0|(\hat S_{\vv{n}_0,+}-\hat S_{\vv{n}_0,-})\lvert \vv{n}_0\rangle=0.
\end{equation}

Therefore, we reach
\begin{equation}
\langle \vv{n}_0\rvert\,\hat{\vv S}\,\lvert \vv{n}_0\rangle
= \hbar\,s\,\vv{n}_0.
\label{eq:S_expect}
\end{equation}
Substituting Eq. \ref{eq:Tr_rho_S} and \(\vv{n}_0=\vv S_{\mathrm{cl}}/(\hbar\,s)\) (Eq. \ref{eq:n_hat_def}), we have proven Eq. \ref{eq:exp_S=S_cl}.

Second, using the von Neumann equation (Eq. \ref{eq:vNEq}) and the cyclic property of the trace, we compute the time derivative:
\begin{equation}
\frac{d}{dt}\Tr\bigl({\rho}_{\vv{n}_0}\,\hat{\vv S}\,\bigr)
=\frac{1}{i \hbar}\,
\Tr\bigl([\hat{H},{\rho}_{\vv{n}_0}]\,\hat{\vv S}\,\bigr)
=\frac{1}{i \hbar}\,
\Tr\bigl({\rho}_{\vv{n}_0}\,[\hat{\vv S},\hat{H}]\,\bigr).
\label{eq:dScl_trace}
\end{equation}
Note that
\begin{equation}
    [\hat{\vv S},\hat{H}]
\;=\;
\bigl([\hat{S}_x,\hat{H}],\,[\hat{S}_y,\hat{H}],\,[\hat{S}_z,\hat{H}]\bigr).
\end{equation}

Substituting the Hamiltonian \(\hat{H}=-\gamma\,\vv B\cdot\hat{\vv S}\), using the commutator identity \([\hat{\vv S},\vv B\cdot\hat{\vv S}]=-i \hbar\,\hat{\vv S}\times\vv B\) (see Appendix \ref{sec:CommCross2}), then inserting Eq. \ref{eq:exp_S=S_cl}, we obtain
\begin{equation}
\frac{d}{dt}\Tr\bigl({\rho}_{\vv{n}_0}\,\hat{\vv S}\,\bigr)
=\frac{1}{i \hbar}\,
\Tr\bigl({\rho}_{\vv{n}_0}\,[\hat{\vv S},-\gamma\,\vv B\cdot\hat{\vv S}\,]\bigr)
=\gamma\,
\Tr\bigl({\rho}_{\vv{n}_0}\,\hat{\vv S}\times\vv B\,\bigr)
=\gamma\,\vv S_{\mathrm{cl}}\times\vv B.
\label{eq:dScl_comm}
\end{equation}
Finally, recognizing \(\vv{n}_0=\vv S_{\mathrm{cl}}/(\hbar\,s)\) (Eq. \ref{eq:n_hat_def}) gives
\begin{equation}
\frac{d\vv{n}_0}{dt}
=\gamma\,\vv{n}_0\times\vv B,
\label{eq:Majorana_Bloch}
\end{equation}
which is the classical Majorana--Bloch equation (Eq. \ref{eq:MBE_n}).  

Alternatively, we may recover the classical Majorana--Bloch equation by applying the Ehrenfest theorem,
\begin{equation}
    \frac{d}{dt}\langle \hat{\mathcal{O}}\rangle
=\frac{1}{i\hbar}\,\langle[\hat{\mathcal{O}},\hat{H}]\rangle,
\end{equation}
to the spin operator, where the general operator $\hat{\mathcal{O}}=\hat{\vv S}$. Again, substituting the Hamiltonian \(\hat{H}=-\gamma\,\vv B\cdot\hat{\vv S}\) and using the commutator identity \([\hat{\vv S},\vv B\cdot\hat{\vv S}\,]=-i \hbar\,\hat{\vv S}\times\vv B\) (see Appendix \ref{sec:CommCross2}), we obtain
\begin{equation}
\frac{d}{dt}\langle\hat{\vv S}\rangle
=\frac{1}{i\hbar}\,\langle[\hat{\vv S},-\gamma\,\vv B\cdot\hat{\vv S}\,]\rangle
=\gamma\,\langle\,\hat{\vv S}\times\vv B\,\rangle
=\gamma\,\langle\hat{\vv S}\rangle\times\vv B.
\end{equation}
Since $\langle\hat{\vv S}\rangle=\hbar\,s\,\vv{n}_0$ (Eq. \ref{eq:S_expect}) for coherent states, we recover exactly the classical Majorana--Bloch equation (Eq. \ref{eq:MBE_n}). Here \(\vv B\) is treated as a classical control field, so it commutes with \(\hat{\vv S}\) and may be taken outside the trace expression. This identity would not hold if \(\vv B\) were promoted to an operator-valued field.

\section{Applicability to coherent states}
The derivation presented above applies to coherent pure states or statistical mixtures thereof only. A coherent state is a particular family of pure states constructed to mimic classical behavior as closely as quantum mechanics allows. 

Any direction in space \(\mathbb{R}^3\) may be specified by the polar angle \(\theta\in[0,\pi]\) and the azimuthal angle \(\phi\in[0,2\pi)\) as
\begin{equation}
\vv{n}_0(\theta,\phi)
= \bigl(\sin\theta\cos\phi,\;\sin\theta\sin\phi,\;\cos\theta\bigr).
\end{equation}
Solving the eigenvalue equation
\begin{equation}
    {\rho}_{\vv{n}_0}\,\lvert\vv{n}_0(\theta,\phi)\rangle = \lvert\vv{n}_0(\theta,\phi)\rangle
\end{equation}
yields the coherent‑state ket in the $\lvert s,m\rangle$ basis with quantization along \(\hat z\) \cite{kam2023coherent, Arecchi1972Atom},
\begin{equation}
    \lvert \vv{n}_0(\theta,\phi)\rangle
=\sum_{m=-s}^{+s}
\sqrt{\binom{2s}{s+m}}\,
e^{i (s-m)\phi}\,
\Bigl(\cos\tfrac{\theta}{2}\Bigr)^{s+m}
\Bigl(\sin\tfrac{\theta}{2}\Bigr)^{\,s-m}
\;\lvert s,m\rangle.
\end{equation}
For \(s=\tfrac12\),
\begin{equation}
\lvert \vv{n}_0\rangle
=\cos\tfrac{\theta}{2}\,\lvert\tfrac12,\tfrac12\rangle
+e^{i\phi}\,\sin\tfrac{\theta}{2}\,\lvert\tfrac12,-\tfrac12\rangle.
\end{equation}
For \(s=1\),
\begin{equation}
\lvert \vv{n}_0\rangle
=\cos^2\tfrac{\theta}{2}\,\lvert1,1\rangle
+\sqrt{2}\,e^{i\phi}\,\sin\tfrac{\theta}{2}\cos\tfrac{\theta}{2}\,\lvert1,0\rangle
+e^{2i\phi}\,\sin^2\tfrac{\theta}{2}\,\lvert1,-1\rangle.
\end{equation}
Note that \(\lvert1,0\rangle\) is a non-coherent pure state because no values of \(\theta, \phi\) yield \(\lvert \vv{n}_0\rangle = \lvert1,0\rangle\). The spin-1 state $\lvert 1, 0 \rangle$ is represented on the Majorana sphere by two antipodal stars, meaning they lie on opposite sides of the sphere. This configuration reflects a state with zero average spin and purely quadrupolar character. In contrast, a coherent state corresponds to two coincident stars at the same point on the sphere, indicating maximal polarization along a single direction.

The set of all spin-\(s\) coherent-state rays 
\begin{equation}
\bigl\{[\,\lvert \vv{n}_0(\theta,\phi)\rangle\,]\colon (\theta,\phi)\in S^{2}\bigr\}
\;\subset\;\mathbb{P}(\mathscr H)
\end{equation}
is homeomorphic to the coset manifold \(\mathrm{SU}(2)/\mathrm{U}(1)\), which is homeomorphic to \(S^{2}\ni(\theta,\phi)\). Here, the square brackets denote the ray in projective Hilbert space associated with the ket---that is, the equivalence class of all vectors differing by a global phase. The homeomorphisms can be written as
\begin{equation}
\bigl\{[\,\lvert \vv{n}_0(\theta,\phi)\rangle\,]\colon(\theta,\phi)\in S^{2}\bigr\}
\;\cong\;
\mathrm{SU}(2)/\mathrm U(1)
\;\cong\;
S^{2}\,.
\end{equation}
Put differently, the SU(2) coherent-state rays form a submanifold of the projective Hilbert space \(\mathbb{P}(\mathscr H)\) for arbitrary spin \(s\), which is homeomorphic to the two-sphere $S^2$, a manifold of real dimension two. By contrast, the full pure-state space of a \((2s+1)\)-level system is the complex projective space \(\mathbb{CP}^{2s}\) with real dimension \(4s\) (complex dimension \(2s\)).  

In the special case \(s=\tfrac12\), one has 
\(\mathbb{CP}^{1}\cong S^{2}\); thus, every pure state is an SU(2) coherent state.  Stated otherwise, for spin \(\tfrac12\), no pure state lies outside the coherent-state manifold, and the Bloch sphere parametrizes all pure states.  

One may extend from coherent states to mixed states that specifically originate from a combination of coherent states:
\begin{equation}
    {\rho}_{\vv{n}_0^{(i)}} \;=\;\bigl|\vv{n}_0^{(i)}\bigr\rangle\!\bigl\langle \vv{n}_0^{(i)}\bigr|
,\quad
{\rho} \;=\;\sum_{i}p_{i}\,{\rho}_{\vv{n}_0^{(i)}},
\end{equation}
subject to the convexity conditions of
\begin{equation}
    \sum_{i}p_{i}=1,\;p_{i}\ge0.
\end{equation}
From Eq. \ref{eq:vNEq} for coherent states, we obtain the associated von Neumann equation for this special set of states:
\begin{equation}
i\hbar\,\frac{d{\rho}}{dt}
=[\hat{H},{\rho}].
\end{equation}

Although the von Neumann equation governs the evolution of any density operator, extending our approach to arbitrary states demands caution. A noncoherent pure state lacks a single classical spin pseudovector representation; except in the trivial pure‐state case, a mixed-state density operator possesses an uncountable continuum of distinct decompositions into pure-state ensembles.

\section{Generalization to arbitrary spin in any state}

For general spin states, we employ spin-$\tfrac{1}{2}$ primitives \cite{sakurai2020modern} and follow the Majorana construction or the Schur--Weyl duality \cite{majorana1932atomi, majorana2020oriented, chase2008collective, hassani2013mathematical}. 
First, the Majorana--Bloch equation for each primitive is converted into the equivalent von Neumann equation for the single-qubit density operator. 
Next, we take tensor products of these primitives and form the full composite density operator. 
Finally, applying the symmetrization projector $\hat P_{\mathrm{sym}}$ to this composite density operator restricts the state to the symmetric subspace, which carries the spin-$s$ representation. 
Therefore, any spin-$s$ state---pure or mixed---can be obtained from spin-$\tfrac{1}{2}$ constituents by symmetrization of their tensor products. At every stage of the conversion, the corresponding von Neumann equation remains valid.

As shown in the earlier section, the classical Majorana--Bloch equation for each primitive \(a\),
\begin{equation}
\frac{d\vv {n}_0^{(a)}}{dt}
=\gamma\,\vv {n}_0^{(a)}\times\vv B,
\label{eq:primitive-bloch-eom}
\end{equation}
can be converted to the von Neumann equation in any pure state for a spin \(s=\tfrac12\). 

Each primitive \(a\) lives on \(\mathbb C^2\) with identity \(\hat I_2\) and Pauli operators \(\hat \sigma^{(a)}_i\), where $i\in\{x,y,z\}$. We write
\begin{equation}
\hat S^{(a)}_i=\frac{\hbar}{2}\,\hat \sigma^{(a)}_i,
\qquad
[\,\hat S^{(a)}_i,\hat S^{(b)}_j\,]=i\hbar\,\delta_{ab}\,\varepsilon_{ijk}\,\hat S^{(a)}_k.
\label{eq:primitive-ops}
\end{equation}
The individual Zeeman-type Hamiltonian is
\begin{equation}
    \hat H^{(a)} = -\,\gamma\,\vv B\cdot\hat{\vv S}^{(a)}.
\end{equation}
For each primitive, we define
\begin{equation}
\hat X^{(a)}_{\vv{n}_0}\equiv\frac{1}{\hbar}\,\vv{n}_0\cdot\hat{\vv S}^{(a)}=\frac{1}{2}\,\vv{n}_0\cdot\hat{\vv{\sigma}}^{(a)}.
\label{eq:primitive-HX}
\end{equation}
Repeating the steps in the earlier section gives
\begin{equation}
\frac{d\hat X^{(a)}_{\vv{n}_0}}{dt}
=\frac{1}{i\hbar}\,\bigl[\,\hat H^{(a)},\hat X^{(a)}_{\vv{n}_0}\,\bigr].
\label{eq:primitive-vNE}
\end{equation}
For \(s=\tfrac12\), the polynomial (Eq. \ref{eq:rho_poly1}) reduces to
\begin{equation}
\mathscr\rho^{(a)}_{\vv{n}_0}
=\hat X^{(a)}_{\vv{n}_0}+\frac{1}{2}\,\hat I_2
=\frac{1}{2}\Bigl(\hat I_2+\vv{n}_0\cdot\hat{\vv{\sigma}}^{(a)}\Bigr).
\label{eq:rho-primitive-coh}
\end{equation}
For coherent states, the von Neumann equation
\begin{equation}
i\hbar\,\frac{d\mathscr\rho^{(a)}_{\vv{n}_0}}{dt}
=\bigl[\,\hat H^{(a)},\mathscr\rho^{(a)}_{\vv{n}_0}\,\bigr].
\label{eq:primitive-vNE-CS}
\end{equation}
Because for \(s=\tfrac12\) all pure states are coherent states, the von Neumann equation for any pure state is obtained:
\begin{equation}
i\hbar\,\frac{d\rho^{(a)}}{dt}
=\bigl[\,\hat H^{(a)},\rho^{(a)}\,\bigr].
\label{eq:primitive-vNE-gen}
\end{equation}

For higher spins, we recapitulate the Majorana construction first for spin one for readability  (see Appendix \ref{app:Majorana-one}), then for arbitrary spin for completeness (see Appendix \ref{app:Majorana-s}). Consequently, we reach the von Neumann equation for spin $s$ in pure states:
\begin{equation}
i\hbar\,\frac{d\rho_{s}}{dt}
=\bigl[\,\hat H_{s},\,\rho_{s}\,\bigr].
\label{eq:vNE-spin-s-main}
\end{equation}

We now extend from pure states to mixed states:
\begin{equation}
{\rho} \;=\;\sum_{i} {p_{i}\,\rho_{s}^i},
\end{equation}
subject to the convexity conditions of
\begin{equation}
    \sum_{i}p_{i}=1,\;p_{i}\ge0.
\end{equation}
From Eq. \ref{eq:vNE-spin-s-main} for pure states, we readily obtain the associated von Neumann equation for any states, including mixed states:
\begin{equation}
i\hbar\,\frac{d{\rho}}{dt}
=[\hat{H_{s}},{\rho}].
\end{equation}

\section{Conclusion}

We have generalized our earlier derivation from the Majorana--Bloch equation to the von Neumann equation for spin-\(\tfrac{1}{2}\) to arbitrary spin-$s$ within the pure coherent-state manifold.
By exploiting the polynomial form of the coherent‐state projector of the spin, we have shown that the time evolution of the projector satisfies the von Neumann equation.  This derivation establishes a bijective correspondence between the classical Majorana--Bloch equation on the real-space sphere and its coherent-state quantum analog in the Hilbert-space operator formalism; therefore, classical and quantum angular-momentum dynamics on the pure coherent-state manifold are in one-to-one correspondence. As expected from the Ehrenfest theorem, taking the expectation value of the angular‐momentum operator in a coherent state exactly reproduces the classical Majorana--Bloch equation. We further generalized the derived von Neumann equation to encompass mixed states that originate from convex combinations of coherent‐state projectors. 

Finally, we demonstrated that the Majorana construction offers a unified approach for building from spin-$\tfrac{1}{2}$ primitives to arbitrary spin-$s$. This framework applies equally to pure and mixed states, thereby providing a systematic description of spin dynamics within the symmetric subspace. The conclusion applies to orbital angular momentum, too.

The derivations remain valid for a time-dependent field 
\(\vv B=\vv B(t)\), since no time derivatives of \(\vv B\) enter the commutator relations. However, the present work is restricted to Hamiltonians that are linear in spin. Within this regime, the bijection is valid on the coherent-state manifold. The mapping does not extend to non-linear spin Hamiltonians.

\section*{Acknowledgments}
We thank our team members---Kelvin Titimbo, Suleyman Kahraman, Xukun Lin, Qihang Liu, Mark Zhu, and Arthur Chang---for their fruitful discussions.

\clearpage

\appendix
\setcounter{secnumdepth}{2}


\section{Proof of the commutation relation for \([\vv a\cdot\hat{\vv S},\;\vv b\cdot\hat{\vv S}\,]\)} \label{sec:CommCross}

We show that for any vectors \(\vv{a}, \vv{b} \in \mathbb{R}^3\),
\begin{equation}
[\vv a\cdot\hat{\vv S},\;\vv b\cdot\hat{\vv S}\,]
=i\hbar\;(\vv a\times\vv b\,)\cdot\hat{\vv S}.
\end{equation}
Substituting
\begin{equation}
\vv a\cdot\hat{\vv S}
=\sum_{i}a_{i}\,\hat{S}_i,
\quad
\vv b\cdot\hat{\vv S}
=\sum_{j}b_{j}\,\hat{S}_j
\end{equation}
into the left side, we reach
\begin{equation}
[\vv a\cdot\hat{\vv S},\;\vv b\cdot\hat{\vv S}\,]
=\sum_{i,j}a_{i}b_{j}\,[\hat{S}_i,\hat{S}_j].
\end{equation}
Resorting to
\begin{equation}
[\hat{S}_i,\hat{S}_j]
= i\hbar \sum_{k=1}^{3}\varepsilon_{ijk}\,\hat{S}_k,
\end{equation}
where $\varepsilon_{ijk}$ is the Levi--Civita symbol, we obtain
\begin{equation}
[\vv a\cdot\hat{\vv S},\;\vv b\cdot\hat{\vv S}\,]
=\sum_{i,j}a_{i}b_{j}\,\biggl(\sum_{k}i\hbar\,\varepsilon_{ijk}\,\hat{S}_k\biggr)
=i\hbar\sum_{i,j,k}a_{i}b_{j}\,\varepsilon_{ijk}\,\hat{S}_k.
\end{equation}
By definition of the cross‐product, one has
\begin{equation}
(\vv a\times\vv b\,)_{k}
=\sum_{i,j}\varepsilon_{ijk}\,a_{i}\,b_{j}.
\end{equation}
Substitution yields
\begin{equation}
[\vv a\cdot\hat{\vv S},\;\vv b\cdot\hat{\vv S}\,]
=i\hbar\sum_{k}(\vv a\times\vv b\,)_{k}\,\hat{S}_k
=i\hbar\;(\vv a\times\vv b\,)\cdot\hat{\vv S}.
\end{equation}

\section{Proof of the density operator}  \label{sec:rho}

\(\hat{X}_{\vv{n}_0}\) is a Hermitian operator on a \((2s+1)\)‑dimensional space with simple (nondegenerate) spectrum
\begin{equation}
\{\,m : m=-s,\,-s+1,\dots,s\}
\end{equation}
and eigenvectors \(\{\lvert m\rangle_{\vv{n}_0}\}\) satisfying
\begin{equation}
\hat{X}_{\vv{n}_0}\,\lvert m\rangle_{\vv{n}_0} = m\,\lvert m\rangle_{\vv{n}_0},
\quad
\langle m \mid m'\rangle_{\vv{n}_0}
= \delta_{m,m'}.
\label{eq:X_eigen}
\end{equation}
We prove (Eq. \ref{eq:rho_poly1})
\begin{equation}
{\rho}_{\vv{n}_0}
= \prod_{m'=-s}^{s-1}
  \frac{\hat{X}_{\vv{n}_0} - m' \hat{I}}{s - m'},
\label{eq:rho_poly}
\end{equation}
which by construction omits the factor with \(m'=s\).  We claim \({\rho}_{\vv{n}_0}\) is the rank‑one projector onto the eigenspace of \(\hat{X}_{\vv{n}_0}\) with eigenvalue \(s\), i.e., the density operator. It is a specific realization of Löwdin’s general projection operator \cite{Crossley1977, Lowdin1955}.

Since each factor \(\bigl(\hat{X}_{\vv{n}_0} - m' \hat{I}\bigr)/(s - m')\) commutes with \(\hat{X}_{\vv{n}_0}\), the operator \({\rho}_{\vv{n}_0}\) is diagonal in the basis \(\{\lvert m\rangle_{\vv{n}_0}\}\). It follows that
\begin{equation}
{\rho}_{\vv{n}_0}\,\lvert m\rangle_{\vv{n}_0}
= \prod_{m'=-s}^{s-1}
  \frac{m - m'}{s - m'}\,\lvert m\rangle_{\vv{n}_0}.
\end{equation}
If \(m = s\), we have
\begin{equation}
{\rho}_{\vv{n}_0}\,\lvert s\rangle_{\vv{n}_0} = \lvert s\rangle_{\vv{n}_0}.
\label{eq:rho_eigen}
\end{equation}
If instead \(m \neq s\), then
\begin{equation}
{\rho}_{\vv{n}_0}\,\lvert m\rangle_{\vv{n}_0} = 0.
\end{equation}
Namely, \({\rho}_{\vv{n}_0}\) acts as the identity on \(\lvert s\rangle_{\vv{n}_0}\) and annihilates all other \(\lvert m\neq s\rangle\).  Therefore, we have
\begin{equation}
{\rho}_{\vv{n}_0} 
= \lvert s\rangle_{\vv{n}_0} \langle s\rvert_{\vv{n}_0},
\label{eq:rho_def}
\end{equation}
which is the unique rank‑one projector onto the eigenspace of \(\hat{X}_{\vv{n}_0}\) with the eigenvalue \(s\) and is hence idempotent (i.e., \( {\rho}_{\vv{n}_0}^2 = {\rho}_{\vv{n}_0} \)).

For $s = \tfrac{1}{2}$, the density operator (Eq. \ref{eq:rho_poly1} or \ref{eq:rho_poly}) truncates at degree one:
\begin{equation}
{\rho}_{\vv{n}_0} = \frac{1}{2}\hat{I} + \frac{1}{\hbar}\,\vv{n}_0\cdot\hat{\vv S},
\end{equation}
which is a pure qubit state. The state points in the direction \( \vv{n}_0 \) on the Bloch sphere.

\section{Proof of the commutator expansion} \label{sec:CommExp}

The identity
\begin{equation}
\sum_{j=0}^{k-1} \hat{X}^j [\hat{X}, \hat{H}] \hat{X}^{k-1-j}
= [\hat{X}^k, \hat{H}]
\end{equation}
for any operators $\hat{X}$ and $\hat{H}$ can be proven by induction.

Step 1. Base case $k = 1$:

We have
\begin{equation}
[\hat{X}, \hat{H}] = [\hat{X}^1, \hat{H}],
\end{equation}
and the sum on the left side becomes
\begin{equation}
\sum_{j=0}^{0} \hat{X}^0 [\hat{X},\hat{H}] \hat{X}^{0} = [\hat{X}^1, \hat{H}].
\end{equation}
Therefore, the identity holds for $k = 1$.

Step 2. Inductive hypothesis:

Assume the identity holds for integer $k \ge 1$:
\begin{equation}
\sum_{j=0}^{k-1} \hat{X}^j [\hat{X}, \hat{H}] \hat{X}^{k-1-j} = [\hat{X}^k, \hat{H}].
\end{equation}

Step 3. Inductive step:

From the product rule,
\begin{equation}
[\hat{A}\hat{B}, \hat{C}] = \hat{A}[\hat{B},\hat{C}] + [\hat{A},\hat{C}]\hat{B},
\end{equation}
we reach
\begin{equation}
[\hat{X}^{k+1}, \hat{H}] = [\hat{X}^k \hat{X}, \hat{H}] = \hat{X}^k [\hat{X}, \hat{H}] + [\hat{X}^k, \hat{H}] \hat{X}.
\label{eq:Xk+1H}
\end{equation}

Applying the inductive hypothesis to $[\hat{X}^k, \hat{H}]$ yields
\begin{equation}
[\hat{X}^k, \hat{H}] \hat{X} = \bigl( \sum_{j=0}^{k-1} \hat{X}^j [\hat{X},\hat{H}] \hat{X}^{k-1-j} \bigr) \hat{X}
= \sum_{j=0}^{k-1} \hat{X}^j [\hat{X},\hat{H}] \hat{X}^{k - j}.
\end{equation}
Substitution into Eq. \ref{eq:Xk+1H} yields
\begin{equation}
[\hat{X}^{k+1}, \hat{H}]
= \hat{X}^k [\hat{X}, \hat{H}] \hat{X}^0 + \sum_{j=0}^{k-1} \hat{X}^j [\hat{X},\hat{H}] \hat{X}^{k - j}
= \sum_{j=0}^{k} \hat{X}^j [\hat{X},\hat{H}] \hat{X}^{k - j}.
\end{equation}
Therefore, the identity holds for $k+1$. By induction, the formula is true for all integers $k \ge 1$.

\section{Proof of the commutation relation for \([\hat{\vv S},\vv B\cdot\hat{\vv S}\,]\)}  \label{sec:CommCross2}

For the $i$th component of \(\hat{\vv S}\), we have
\begin{equation}
\bigl[\hat{S}_i,\; \vv{B} \cdot \hat{\vv S} \,\bigr]
= \bigl[ \hat{S}_i,\; \sum_{j} B_j \hat{S}_j \bigr]
= \sum_{j} B_j [\hat{S}_i, \hat{S}_j].
\label{eq:Si_BS}
\end{equation}
The spin operators obey
\begin{equation}
\bigl[\hat{S}_i,\,\hat{S}_j\bigr] = i\hbar\,\sum_{k} \varepsilon_{ijk}\,\hat{S}_k,
\end{equation}
where $\varepsilon_{ijk}$ is the Levi--Civita symbol. Substituting into Eq. \ref{eq:Si_BS} yields
\begin{equation}
\bigl[ \hat{S}_i,\; \vv{B}\cdot\hat{\vv S}\,\bigr]
= \sum_{j}B_j\,[\hat{S}_i,\hat{S}_j]
= i\hbar\,\sum_{j,k}\varepsilon_{ijk}\,B_j\,\hat{S}_k
= i\hbar\, \bigl( \vv{B} \times \hat{\vv S} \bigr)_i.
\end{equation}
Thus, we reach
\begin{equation}
\bigl[\hat{\vv S},\, \vv{B} \cdot \hat{\vv S}\,\bigr] 
= i\hbar\, \vv{B} \times \hat{\vv S} 
= -\,i\hbar\,\hat{\vv S}\times\vv B.
\end{equation}

\section{Majorana construction for spin one in pure state} \label{app:Majorana-one}

\subsection{Tensor-product space}

We consider two spin-\(\tfrac{1}{2}\) constituents labeled by \(a\in\{1,2\}\) and assume \(\gamma\) and \(\vv B\) are independent of \(a\) so that both spins couple identically to the external field. Let \(\rho\) be a density operator on \(\mathbb C^2\otimes\mathbb C^2\), and for concreteness, one may start from a product
\begin{equation}
    \rho\equiv\rho^{(1)}\otimes\rho^{(2)}.
\end{equation}
The individual Hamiltonian is
\begin{equation}
    \hat H^{(a)}=-\gamma\,\vv B\cdot\hat{\vv S}^{(a)}.
\end{equation}
The two-spin Hamiltonian is
\begin{equation}
\hat H_{\mathrm{tot}}\equiv\hat H^{(1)}\otimes \hat I_2+\hat I_2\otimes\hat H^{(2)}
=-\,\gamma\,\vv B\cdot\Bigl(\hat{\vv S}^{(1)}+\hat{\vv S}^{(2)}\Bigr).
\label{eq:Htot}
\end{equation}
The von Neumann equation for the two-spin system is
\begin{equation}
i\hbar\,\frac{d\rho}{dt}=\bigl[\,\hat H_{\mathrm{tot}},\rho\,\bigr].
\label{eq:vNE-two}
\end{equation}

\subsection{Symmetrization}
We introduce the permutation operator \(\hat{P}_\pi\).  Acting on the product basis states of two spin-\(\tfrac{1}{2}\) particles yields
\begin{equation}
    \hat{P}_\pi \ket{\uparrow \uparrow} = \ket{\uparrow \uparrow},\quad
\hat{P}_\pi \ket{\uparrow \downarrow} = \ket{\downarrow \uparrow},\quad
\hat{P}_\pi \ket{\downarrow \uparrow} = \ket{\uparrow \downarrow},\quad
\hat{P}_\pi \ket{\downarrow \downarrow} = \ket{\downarrow \downarrow}.
\end{equation}

We then define the symmetric projector
\begin{equation}
\hat P_{\mathrm{sym}}\equiv\frac{\hat I_4+\hat P_\pi}{2}.
\label{eq:Psym-Pasym}
\end{equation}
For comparison, the antisymmetric projector is \(\hat P_{\mathrm{asym}}\equiv\tfrac{1}{2}\bigl(\hat I_4-\hat P_\pi\bigr)\).

Permutation invariance gives
\begin{equation}
\bigl[\,\hat H_{\mathrm{tot}},\hat P_\pi\,\bigr]=0,
\qquad
\bigl[\,\hat H_{\mathrm{tot}},\hat  P_{\mathrm{sym}}\,\bigr]=0.
\label{eq:perm-inv}
\end{equation}
Thus, the triplet sector is dynamically isolated from the singlet one.

We next define the spin-one density operator as the symmetric-sector restriction.
\begin{equation}
\rho_{(1)}\equiv\hat P_{\mathrm{sym}}\,\rho\,\hat P_{\mathrm{sym}}.
\label{eq:rho1-def}
\end{equation}
One can show \(\operatorname{Tr} \rho_{(1)} = \operatorname{Tr} \left( \hat{P}_{\mathrm{sym}} \rho \right)\) using cyclicity of trace and idempotency of \(\hat{P}_{\mathrm{sym}}\).

As a \(4\times4\) operator, \(\rho_{(1)}\) has support only on the symmetric sector; it acts effectively on a three-dimensional subspace. We note that \(\rho_{(1)}\) is positive semidefinite:
\begin{equation}
\langle v|\rho_{(1)}|v\rangle
=\langle v|\,\hat P_{\mathrm{sym}}\rho\,\hat P_{\mathrm{sym}}|v\rangle
=\langle w|\rho|w\rangle\ge 0,
\qquad
|w\rangle=\hat P_{\mathrm{sym}}|v\rangle.
\end{equation}
The symmetric projector produces the entire triplet subspace, of which the spin-one coherent states form a subset where the two primitive spin-\(\tfrac{1}{2}\) states are aligned.

We differentiate \(\rho_{(1)}=\hat P_{\mathrm{sym}}\,\rho\,\hat P_{\mathrm{sym}}\) using the time independence of \(\hat P_{\mathrm{sym}}\)
\begin{equation}
\frac{d\rho_{(1)}}{dt}=\hat P_{\mathrm{sym}}\frac{d\rho}{dt}\hat P_{\mathrm{sym}}.
\label{eq:drho1dt-step1}
\end{equation}
We insert \eqref{eq:vNE-two} and expand the commutator
\begin{equation}
i\hbar\,\frac{d\rho_{(1)}}{dt}
=\hat P_{\mathrm{sym}}\hat H_{\mathrm{tot}}\rho\,\hat P_{\mathrm{sym}}
-\hat P_{\mathrm{sym}}\rho\,\hat H_{\mathrm{tot}}\hat P_{\mathrm{sym}}.
\label{eq:drho1dt-step2}
\end{equation}
Using \([\hat H_{\mathrm{tot}},\hat P_{\mathrm{sym}}]=0\) and \(\rho_{(1)}=\hat P_{\mathrm{sym}}\rho\,\hat P_{\mathrm{sym}}\) yields
\begin{equation}
i\hbar\,\frac{d\rho_{(1)}}{dt}
=\Bigl[\,\hat P_{\mathrm{sym}}\hat H_{\mathrm{tot}}\hat P_{\mathrm{sym}},\ \rho_{(1)}\,\Bigr].
\label{eq:vNE-spin1}
\end{equation}
We define the projected spin-one Hamiltonian
\begin{equation}
\hat H_{(1)}
\equiv \hat P_{\mathrm{sym}}\hat H_{\mathrm{tot}}\hat P_{\mathrm{sym}}.
\label{eq:spin1-H}
\end{equation}
We define the projected total spin operator,
\begin{equation}
\hat{\vv S}\big|_{s=1}
\equiv \hat P_{\mathrm{sym}}\,
\bigl(\hat{\vv S}^{(1)}+\hat{\vv S}^{(2)}\bigr)\,
\hat P_{\mathrm{sym}},
\label{eq:spin1-S}
\end{equation}
leading to
\begin{equation}
    \hat H_{(1)}=-\gamma\,\vv B\cdot\hat{\vv S}\big|_{s=1}.
\end{equation}

Therefore, the induced dynamics close within the triplet sector
\begin{equation}
i\hbar\,\frac{d\rho_{(1)}}{dt}
= [\,\hat H_{(1)},\rho_{(1)}\,].
\label{eq:vNE-spin1-clean}
\end{equation}
If \(\rho\) is initially supported entirely on the triplet, then \(\rho_{(1)}=\rho\) and \(\operatorname{Tr}\rho_{(1)}=1\). Otherwise, \(\operatorname{Tr}\rho_{(1)}\) is the conserved triplet population, and a normalized spin-one state is \(\varrho_{(1)}\equiv \rho_{(1)}/\operatorname{Tr}\rho_{(1)}\).

\subsection{Coupled-basis representation}
We reorganize the tensor-product space into irreducible components of total spin via the Clebsch–Gordan transform. We change from the uncoupled basis to the coupled basis
\begin{equation}
\bigl\{|1,1\rangle,\ |1,0\rangle,\ |1,-1\rangle,\ |0,0\rangle\bigr\}
\label{eq:coupled-basis}
\end{equation}
with
\begin{equation}
|1,1\rangle=\ket{\uparrow\uparrow},\quad
|1,0\rangle=\frac{\ket{\uparrow\downarrow}+\ket{\downarrow\uparrow}}{\sqrt2},\quad
|1,-1\rangle=\ket{\downarrow\downarrow},\quad
|0,0\rangle=\frac{\ket{\uparrow\downarrow}-\ket{\downarrow\uparrow}}{\sqrt2}.
\label{eq:CG-kets}
\end{equation}
Let \(U\) be the Clebsch–Gordan unitary whose columns are the vectors in \eqref{eq:CG-kets} written in the uncoupled basis \(\{\ket{\uparrow\uparrow},\ket{\uparrow\downarrow},\ket{\downarrow\uparrow},\ket{\downarrow\downarrow}\}\):
\begin{equation}
U=\begin{pmatrix}
1&0&0&0\\
0&\tfrac1{\sqrt2}&0&\tfrac1{\sqrt2}\\
0&\tfrac1{\sqrt2}&0&-\tfrac1{\sqrt2}\\
0&0&1&0
\end{pmatrix},
\qquad
U^\dagger U=I_4.
\label{eq:U-matrix}
\end{equation}

Using \(\hat S_i^{(a)}=\tfrac\hbar2\,\hat{\sigma}_i^{(a)}\), where $i\in\{x,y,z\}$, we reach
\begin{equation}
\hat S_i=\hat S_i^{(1)}+\hat S_i^{(2)}
=\tfrac\hbar2\bigl(\hat{\sigma}_i\otimes \hat I_2+\hat I_2\otimes\hat{\sigma}_i\bigr).
\label{eq:total-Si}
\end{equation}
Transforming with \(U\) yields
\begin{equation}
U^\dagger\,\hat S_i\,U
=
\begin{pmatrix}
\hat{J}_i & 0\\
0 & 0
\end{pmatrix}.
\label{eq:blockdiag}
\end{equation}
The commutators satisfy \([\hat{J}_i,\hat{J}_j]=i\hbar\,\varepsilon_{ijk}\hat{J}_k\), and the Casimir gives \(\hat{J}^2=\hat{J}_x^2+\hat{J}_y^2+\hat{J}_z^2=2\hbar^2\,\hat I_3\). Thus, the upper \(3\times3\) block \(\hat{J}_i\) acts on the triplet, and the lower entry annihilates the singlet. 

Equivalently,
\begin{equation}
U^\dagger\,\hat H_{\mathrm{tot}}\,U
=
\begin{pmatrix}
\hat H_{\mathrm{triplet}} & 0\\
0 & 0
\end{pmatrix},
\label{eq:H-block-triplet}
\end{equation}
where the triplet Hamiltonian
\begin{equation}
\hat H_{\mathrm{triplet}}
\equiv -\gamma\,\vv B\cdot\hat{\vv{J}}.
\label{eq:Htriplet-def}
\end{equation}

In the uncoupled basis, we have
\begin{equation}
 P_\pi=
\begin{pmatrix}
1&0&0&0\\
0&0&1&0\\
0&1&0&0\\
0&0&0&1
\end{pmatrix},
\qquad
 P_{\mathrm{sym}}
=\frac{ I_4+ P_\pi}{2}
=\frac12
\begin{pmatrix}
2&0&0&0\\
0&1&1&0\\
0&1&1&0\\
0&0&0&2
\end{pmatrix}.
\label{eq:Ppi-Psym-explicit}
\end{equation}
In the coupled basis, we obtain
\begin{equation}
\Pi_{\mathrm{sym}} \equiv
U^\dagger\, P_{\mathrm{sym}}\,U
=\mathrm{diag}( I_3,\,0).
\label{eq:Pproj-coupled}
\end{equation}
Thus, \( P_{\mathrm{sym}}\) acts as identity on the triplet and as zero on the singlet. Similarly, 
\begin{equation}
    \Pi_{\mathrm{asym}} \equiv U^\dagger\, P_{\mathrm{asym}}\,U
=\mathrm{diag}(0_3,\,1).
\end{equation}

In the coupled basis \(I_4=\Pi_{\mathrm{sym}}+\Pi_{\mathrm{asym}}\) with \(\Pi_{\mathrm{sym}}\Pi_{\mathrm{asym}}=0\), one can show
\begin{equation}
U^\dagger \rho\, U
=
\begin{pmatrix}
\rho_{\mathrm{triplet}} & C\\
C^\dagger & \rho_{\mathrm{singlet}}
\end{pmatrix},
\qquad
\rho_{\mathrm{triplet}}\in\mathbb C^{3\times3},\ 
C\in\mathbb C^{3\times1},\ 
\rho_{\mathrm{singlet}}\in\mathbb C.
\label{eq:rho-block-general}
\end{equation}

We apply \(U\) to the projected von Neumann equation \eqref{eq:vNE-spin1-clean}. Since \(U\) is time independent, 
\begin{equation}
U^\dagger\!\left(i\hbar\,\frac{d\rho_{(1)}}{dt}\right)\!U
=i\hbar\,\frac{d}{dt}\bigl(U^\dagger\rho_{(1)}U\bigr)
=\bigl[\,U^\dagger\hat H_{(1)}U,\ U^\dagger\rho_{(1)}U\,\bigr].
\end{equation}
We set \(\tilde\rho_{(1)}\equiv U^\dagger\rho_{(1)}U\) and \(\tilde H_{(1)}\equiv U^\dagger\hat H_{(1)}U\), which gives
\begin{equation}
i\hbar\,\frac{d\tilde\rho_{(1)}}{dt}
=\bigl[\,\tilde H_{(1)},\ \tilde\rho_{(1)}\,\bigr].
\end{equation}
For any operator \(A\), one has 
\begin{equation}
U^\dagger \hat P_{\mathrm{sym}} A \hat P_{\mathrm{sym}} U
=\Pi_{\mathrm{sym}}\,(U^\dagger A U)\,\Pi_{\mathrm{sym}}.
\end{equation}
Hence
\begin{equation}
\tilde\rho_{(1)}
=\Pi_{\mathrm{sym}}\,(U^\dagger\rho\,U)\,\Pi_{\mathrm{sym}}
=
\begin{pmatrix}
\rho_{\mathrm{triplet}} & 0\\
0 & 0
\end{pmatrix},
\end{equation}
\begin{equation}
\tilde H_{(1)}
=\Pi_{\mathrm{sym}}\,(U^\dagger\hat H_{\mathrm{tot}}U)\,\Pi_{\mathrm{sym}}
=
\begin{pmatrix}
\hat H_{\mathrm{triplet}} & 0\\
0 & 0
\end{pmatrix}.
\end{equation}

Reading off the triplet block yields the Dicke-basis form
\begin{equation}
i\hbar\,\frac{d\rho_{\mathrm{triplet}}}{dt}
=\bigl[\,\hat H_{\mathrm{triplet}},\ \rho_{\mathrm{triplet}}\,\bigr],
\end{equation}
which is unitarily equivalent to \eqref{eq:vNE-spin1-clean}. Here \(U\) effects only a change of basis on the full space, while the restriction to the symmetric subspace is effected by \(\hat P_{\mathrm{sym}}\). Equivalently, one may first transform \eqref{eq:vNE-two} by \(U\) and then project; both routes give the same triplet dynamics. Since we restrict to the symmetric subspace with total spin one, the working basis is the Dicke basis.

\section{Majorana construction for spin $s$ in pure state} \label{app:Majorana-s}

\subsection{Tensor-product space}

Let $N=2s$ and consider $N$ spin-$\tfrac{1}{2}$ constituents labeled by $a=1,\dots,N$. Assume $\gamma$ and $\vv B$ are independent of $a$ (identical coupling). Let $\rho$ be a density operator on $(\mathbb C^2)^{\otimes N}$; for concreteness, one may start from a product $\rho=\bigotimes_{a=1}^{N}\rho^{(a)}$. The single-spin Hamiltonian is
\begin{equation}
\hat H^{(a)}=-\gamma\,\vv B\cdot\hat{\vv S}^{(a)}.
\end{equation}
The total Hamiltonian and the von Neumann equation are
\begin{equation}
\hat H_{\mathrm{tot}}=-\,\gamma\,\vv B\cdot\sum_{a=1}^{N}\hat{\vv S}^{(a)},
\qquad
i\hbar\,\frac{d\rho}{dt}=\bigl[\,\hat H_{\mathrm{tot}},\rho\,\bigr].
\label{eq:vNE-N}
\end{equation}

\subsection{Symmetrization}

Let $\hat P_\pi$ be the unitary representation of a permutation $\pi\in S_N$. Since $\hat H_{\mathrm{tot}}$ is permutation invariant,
\begin{equation}
\hat P_\pi\,\hat H_{\mathrm{tot}}\,\hat P_\pi^\dagger=\hat H_{\mathrm{tot}}
\quad\Rightarrow\quad
\bigl[\,\hat H_{\mathrm{tot}},\hat P_\pi\,\bigr]=0\ \ \forall\,\pi.
\end{equation}
Let $\hat P_{\mathrm{sym}}^{(N)}$ denote the projector onto the fully symmetric subspace $\mathrm{Sym}^N(\mathbb C^2)$ (dimension $N+1=2s+1$). Define the projected density operator
\begin{equation}
\rho_{(s)}\equiv \hat P_{\mathrm{sym}}^{(N)}\,\rho\,\hat P_{\mathrm{sym}}^{(N)}.
\label{eq:rho-s-def}
\end{equation}
By idempotency and cyclicity, $\operatorname{Tr}\rho_{(s)}=\operatorname{Tr}\!\bigl(\hat P_{\mathrm{sym}}^{(N)}\rho\bigr)$, which is conserved because $[\hat H_{\mathrm{tot}},\hat P_{\mathrm{sym}}^{(N)}]=0$. Thus $\rho_{(s)}$ evolves unitarily within the symmetric sector, and $\varrho_{(s)}\equiv \rho_{(s)}/\operatorname{Tr}\rho_{(s)}$ is the normalized spin-$s$ state.
We note that \(\rho_{(s)}\) is positive semidefinite: \(\langle v \vert \rho_{(s)} \vert v \rangle = \langle w \vert \rho \vert w \rangle, \ket{w} = \hat P_{\mathrm{sym}}^{(N)} \ket{v}\). 
Moreover, \(\Tr \rho_{(s)} = \Tr\!\bigl(\hat P_{\mathrm{sym}}^{(N)}\,\rho\bigr)\) is conserved because \([\hat H_{\mathrm{tot}},\, \hat P_{\mathrm{sym}}^{(N)}]=0\).

Differentiating $\rho_{(s)}$ and using time independence of $\hat P_{\mathrm{sym}}^{(N)}$ gives
\begin{equation}
i\hbar\,\frac{d\rho_{(s)}}{dt}
=\Bigl[\,\hat P_{\mathrm{sym}}^{(N)}\hat H_{\mathrm{tot}}\hat P_{\mathrm{sym}}^{(N)},\ \rho_{(s)}\,\Bigr].
\label{eq:vNE-s}
\end{equation}
Introduce the projected generators and Hamiltonian
\begin{equation}
\hat{\vv S}\big|_{s}\equiv\hat P_{\mathrm{sym}}^{(N)}\Bigl(\sum_{a=1}^{N}\hat{\vv S}^{(a)}\Bigr)\hat P_{\mathrm{sym}}^{(N)},
\qquad
\hat H_{(s)}\equiv-\gamma\,\vv B\cdot\hat{\vv S}\big|_{s},
\label{eq:spin-s-H}
\end{equation}
to obtain the closed spin-$s$ evolution
\begin{equation}
i\hbar\,\frac{d\rho_{(s)}}{dt}
=\bigl[\,\hat H_{(s)},\rho_{(s)}\,\bigr].
\label{eq:vNE-s-clean}
\end{equation}

\subsection{Coupled-basis representation}

Writing single-primitive operators as $\hat S_i^{(a)}=\tfrac{\hbar}{2}\hat{\sigma}_i^{(a)}$ with $i\in\{x,y,z\}$,
\begin{equation}
\hat S_i=\sum_{a=1}^{N}\hat S_i^{(a)}
=\frac{\hbar}{2}\sum_{a=1}^{N}\Bigl(\underbrace{\hat I_2\otimes\cdots\otimes\hat{\sigma}_i}_{a\text{th}}\otimes\cdots\otimes\hat I_2\Bigr).
\end{equation}
Projecting to the symmetric sector defines
\begin{equation}
\hat{\vv S}\big|_{s}
\equiv \hat P_{\mathrm{sym}}^{(N)}\,\hat{\vv S}_{\mathrm{tot}}\,\hat P_{\mathrm{sym}}^{(N)},
\end{equation}
which is unitarily equivalent to the standard $(2s{+}1)\times(2s{+}1)$ representation $\hat{\vv{J}}^{(s)}$. In a coupled basis that orders the fully symmetric spin-$s$ sector first, there exists a unitary $U_N$ such that
\begin{equation}
U_N^\dagger\,\hat P_{\mathrm{sym}}^{(N)}\,U_N=\mathrm{diag}(\hat I_{2s+1},0),
\qquad
U_N^\dagger\,\hat{\vv S}_{\mathrm{tot}}\,U_N=
\begin{pmatrix}
\hat{\vv{J}}^{(s)} & 0\\
0 & \star
\end{pmatrix}.
\label{eq:block-s}
\end{equation}
Here, $\star$ denotes the direct sum \(\bigoplus_{j<s}(\,\hat{\vv J}^{(j)}\otimes \hat I_{m_j}\,)\) taken over the lower-spin irreps with multiplicities \(m_j\). Consequently, for the \emph{projected} Hamiltonian,
\begin{equation}
U_N^\dagger\,\hat H_{(s)}\,U_N=
\begin{pmatrix}
-\gamma\,\vv B\cdot\hat{\vv{J}}^{(s)} & 0\\
0 & 0
\end{pmatrix}.
\label{eq:H-tilde-s}
\end{equation}

Define the coupled-basis operators
\begin{equation}
\tilde\rho_{(s)}\equiv U_N^\dagger\rho_{(s)}U_N,
\qquad
\tilde{\hat H}_{(s)}\equiv U_N^\dagger\hat H_{(s)}U_N.
\label{eq:tildes-s}
\end{equation}
Since $U_N$ is time independent, the von Neumann equation is invariant:
\begin{equation}
i\hbar\,\frac{d\tilde\rho_{(s)}}{dt}
=\bigl[\,\tilde{\hat H}_{(s)},\,\tilde\rho_{(s)}\,\bigr].
\label{eq:vNE-tilde-s}
\end{equation}
Using \eqref{eq:block-s}–\eqref{eq:tildes-s}, the block forms are
\begin{equation}
\tilde\rho_{(s)}=
\begin{pmatrix}
\rho_{s} & 0\\
0 & 0
\end{pmatrix},
\qquad
\tilde{\hat H}_{(s)}=
\begin{pmatrix}
-\gamma\,\vv B\cdot\hat{\vv{J}}^{(s)} & 0\\
0 & 0
\end{pmatrix},
\qquad
\rho_{s}\in\mathbb C^{(2s+1)\times(2s+1)}.
\label{eq:block-forms}
\end{equation}
Reading off the $(2s{+}1)$ block yields the closed spin-$s$ evolution
\begin{equation}
i\hbar\,\frac{d\rho_{s}}{dt}
=\bigl[\,-\gamma\,\vv B\cdot\hat{\vv{J}}^{(s)},\,\rho_{s}\,\bigr].
\label{eq:vNE-final-spin-s}
\end{equation}
Equivalently, defining the coupled-basis spin-$s$ Hamiltonian
\begin{equation}
\hat H_{s}\equiv -\gamma\,\vv B\cdot\hat{\vv{J}}^{(s)},
\label{eq:Hs-def}
\end{equation}
we have
\begin{equation}
i\hbar\,\frac{d\rho_{s}}{dt}
=\bigl[\,\hat H_{s},\,\rho_{s}\,\bigr].
\label{eq:vNE-final-spin-s-H}
\end{equation}

\subsection{Verification}

We now present a proof that symmetrization of $2s$ spin-$\tfrac{1}{2}$ primitives yields the spin-$s$ irreducible representation, demonstrating how the fully symmetric subspace realizes the $\mathrm{SU}(2)$ action.

We take \(N=2s\). The full Hilbert space is \((\mathbb C^2)^{\otimes N}\), spanned by product states of \(N\) spin-\(\tfrac{1}{2}\) primitives. We define the total spin operators
\begin{equation}
\hat S_i \equiv \sum_{a=1}^{N} \hat S_i^{(a)},\qquad
\hat S_{\pm} \equiv \hat S_x \pm i\,\hat S_y ,
\end{equation}
where \(\hat S_i^{(a)}\) acts nontrivially only on the \(a\)-th primitive. On the full space, they satisfy the \(\mathfrak{su}(2)\) commutators
\begin{equation}
[\,\hat S_i,\hat S_j\,]=i\hbar\,\varepsilon_{ijk}\,\hat S_k .
\end{equation}

We first note permutation invariance. For any permutation unitary \(\hat P_\pi\in S_N\),
\begin{equation}
\hat P_\pi\,\hat S_i\,\hat P_\pi^\dagger
=\sum_{a=1}^{N}\hat S_i^{(\pi(a))}=\hat S_i .
\end{equation}
Hence, the total spin operators commute with every permutation. Consequently,
\([\,\hat P_{\mathrm{sym}}^{(N)},\hat S_i\,]=0\), and the symmetric subspace is invariant under the total spin action. That is, if a quantum state lies in the symmetric subspace, then the action of $\hat S_i$ preserves that symmetry and yields another state within the same subspace.

We then construct a highest-weight state. The fully polarized state
\begin{equation}
\ket{\Omega} \equiv \ket{\uparrow}^{\otimes N}\in \mathrm{Sym}^{N}(\mathbb C^{2})
\end{equation}
is symmetric under permutations and satisfies
\begin{equation}
\hat S_z \ket{\Omega} = s\hbar\,\ket{\Omega}, \qquad
\hat S_{+}\ket{\Omega}=0 .
\end{equation}
Hence, \(\ket{\Omega}\) is the highest-weight vector of weight \(s\), i.e., \(\ket{s, s}\).

We build the ladder by repeated application of $\hat S_-$
\begin{equation}
\ket{s,m=s-k}\ \equiv\ C_k\,\hat S_{-}^{k}\ket{\Omega},
\qquad
C_k \equiv \hbar^{-k}\Bigl(\frac{k!\,(2s)!}{(2s-k)!}\Bigr)^{-1/2}.
\end{equation}
The properly normalized ladder state is obtained by applying the lowering operator $k$ times to the highest-weight state $\ket{s, s}$ and compensating for the combinatorial growth of amplitudes with a normalization factor \(C_k\). Because \(\hat S_{-}\) is permutation symmetric and \(\ket{\Omega}\) is symmetric, each \(\ket{s,m}\) lies in \(\mathrm{Sym}^{N}(\mathbb C^{2})\).

We next verify the generator action by using the $\mathfrak{su}(2)$ commutators and the Casimir relation, which shows how the operators preserve the ladder structure:
\begin{equation}
\hat S_z \ket{s,m}= m\hbar\,\ket{s,m}, \qquad
\hat S_{\pm}\ket{s,m}
=\hbar\,\sqrt{s(s+1)-m(m\pm1)}\ \ket{s,m\pm1}.
\end{equation}
Therefore, the set \(\{\ket{s,m}\}_{m=-s}^{s}\) indeed forms a closed \(\mathfrak{su}(2)\) ladder inside the symmetric subspace.


We count the dimension. There are \(N+1=2s+1\) distinct ladder states with different \(\hat S_z\) eigenvalues, hence they are linearly independent. Since \(\dim \mathrm{Sym}^{N}(\mathbb C^{2})=N+1\), these vectors form a basis of the symmetric subspace. Therefore, the symmetric subspace carries an irreducible representation of \(\mathfrak{su}(2)\) with highest weight \(s\).

We finally confirm the Casimir eigenvalue. With
\begin{equation}
\hat S^{2}=\hat S_x^2+\hat S_y^2+\hat S_z^2,
\end{equation}
we have on the highest-weight vector
\begin{equation}
\hat S^{2}\ket{\Omega}=\hbar^{2}s(s+1)\ket{\Omega}.
\end{equation}
Since $\hat S^2$ commutes with all generators and with $\hat S_-$,
\begin{equation}
\hat S^2\bigl(\hat S_-^k\ket{\Omega}\bigr)
= \hat S_-^k \,\hat S^2\ket{\Omega}
= \hbar^2 s(s+1)\,\hat S_-^k\ket{\Omega}.
\end{equation}
Thus, every descendant \(\ket{s,m}\propto\hat S_-^k\ket{\Omega}\) carries the same eigenvalue. On the entire symmetric subspace,
\begin{equation}
\hat S^{2}\big|_{\mathrm{Sym}^{N}}
= \hbar^{2}s(s+1)\,\hat I_{2s+1}.
\end{equation}

We conclude that the restricted operators
\begin{equation}
\hat S_i\big|_{s}\ \equiv\ \hat P_{\mathrm{sym}}^{(N)}\,\hat S_i\,\hat P_{\mathrm{sym}}^{(N)}
\end{equation}
realize the unique \((2s+1)\)-dimensional irrep of \(\mathfrak{su}(2)\).

\section*{References}

\bibliographystyle{apsrev4-2}
\bibliography{refs}

\end{document}